\documentclass{aa}
\usepackage{pstricks}
\usepackage[]{natbib,amsmath,amssymb}
\usepackage{graphicx}
\bibpunct{(}{)}{,}{a}{}{,}

\def\RXJ{{\object{RX J1347.5$-$1145}}}
\def\arcsecf{\!\!^{\prime\prime}}
\def\arcminf{\!\!^{\prime}}

\begin{document}
\title{Strong and weak lensing united I: the combined strong and weak
  lensing cluster mass reconstruction method}
\titlerunning{The combined strong and weak
  lensing cluster mass reconstruction method}
\author{M. Brada\v{c} \inst{1,2,3} \and P. Schneider \inst{1} \and
  M. Lombardi \inst{1,4,5} \and T. Erben \inst{1}}
\offprints{Maru\v{s}a Brada\v{c}}
\mail{marusa@astro.uni-bonn.de}
\institute{Institut f\"ur Astrophysik und Extraterrestrische
  Forschung, Auf dem H\"ugel 71, D-53121 Bonn, Germany
  \and Max-Planck-Institut f\"ur Radioastronomie, Auf dem
  H\"ugel 69, D-53121 Bonn, Germany \and KIPAC, Stanford University, 2575 Sand Hill Road, Menlo Park, CA 94025, USA
  \and European Southern Observatory, Karl-Schwarzschild-Str. 2,
  D-85748 Garching bei M\"unchen, Germany
\and Universit\`{a} degli Studi di Milano,  v. Celoria 16, I-20133
Milano, Italy  
}
\date{Received 22 October 2004 / Accepted 3 March 2005}
%
%

\abstract{ Weak gravitational lensing is considered to be one of the
  most powerful tools to study the mass and the mass distribution of
  galaxy clusters. However, the mass-sheet degeneracy transformation
  has limited its success. We present a novel method for a cluster
  mass reconstruction which combines weak and strong lensing
  information on common scales and can, as a consequence, break the
  mass-sheet degeneracy.  We extend the weak lensing formalism to the
  inner parts of the cluster and combine it with the constraints from
  multiple image systems. We demonstrate the feasibility of the method
  with simulations, finding an excellent agreement between the input
  and reconstructed mass also on scales within and beyond the Einstein
  radius. Using a single multiple image system and photometric
  redshift information of the background sources used for weak and
  strong lensing analysis, we find that we are effectively able to
  break the mass-sheet degeneracy, therefore removing one of the
  main limitations on cluster mass estimates. We conclude that with
  high resolution (e.g. HST) imaging data the method can more
  accurately reconstruct cluster masses and their profiles than
  currently existing lensing techniques.  }

\maketitle
%
%

\def\diff{\mathrm{d}}
\def\ngx{N_{\mathrm{x}}}
\def\ngy{N_{\mathrm{y}}}
\def\eck#1{\left\lbrack #1 \right\rbrack}
\def\eckk#1{\bigl[ #1 \bigr]}
\def\rund#1{\left( #1 \right)}
\def\abs#1{\left\vert #1 \right\vert}
\def\wave#1{\left\lbrace #1 \right\rbrace}
\def\ave#1{\left\langle #1 \right\rangle}

\def\ga#1{\mathcal{G}^{(1)}_{#1}}
\def\gb#1{\mathcal{G}^{(2)}_{#1}}
\def\aa#1{\mathcal{D}^{(1)}_{#1}}
\def\ab#1{\mathcal{D}^{(2)}_{#1}}
\def\k#1{\mathcal{K}_{#1}}
\def\arcsecf{\!\!^{\prime\prime}}
\def\arcminf{\!\!^{\prime}}
\defcitealias{bradac04b}{Paper II}

%
%
\section{Introduction}
\label{sec:introduction}

Clusters of galaxies have long been recognised as excellent
laboratories for many cosmological applications. An especially
important diagnostic is their number density as a function of mass and
redshift. This can only be measured if reliable mass estimates of the
observed clusters can be obtained. In addition, in the framework of
the $\Lambda$CDM cosmological model, the dark matter distribution in
clusters likely follows the NFW profile \citep{navarro97}.

Weak gravitational lensing is one of the most powerful tools currently
available for studying the mass distribution of clusters of
galaxies. The first weak lensing detection in clusters has been made
by \citet{tyson90}.  However, it was only after the pioneering work by
\citet{kaiser93} that the field began to flourish, and since then many
cluster mass reconstructions have been carried out (see
e.g. \citealt{clowe01, clowe02, gavazzi04, lombardi05}). A
disagreement occurring in some cases between the cluster mass
estimated from the weak/strong lensing measurements with X-rays is
still not well understood, although several scenarios have been
proposed to resolve this issue (see e.g.  \citealp{allen98,ettori03}).

In the absence of the redshift information, the main limitation for a
precise weak lensing mass estimate is the mass-sheet degeneracy
\citep{seitz95}. If the redshifts of background sources and/or lens
are not known, the transformation of the surface mass-density $\kappa
\to \kappa' = \lambda \kappa + (1-\lambda)$, where $\lambda \ne 0$ is
an arbitrary constant, leaves the expectation value of measured image
ellipticities unchanged. In \citet{bradac03b} we show that this
degeneracy can be lifted using information on {\it individual} source
redshifts, however only if the weak lensing reconstruction is extended
to the critical parts of the cluster.  Strong lensing is affected by
this transformation as well. Namely, the mass-sheet degeneracy does
not change the image positions (since the source position is not an
observable) and flux ratios and therefore can not be broken if a
single redshift multiple-image system is used. The mass-sheet
degeneracy can in principle be broken using magnification effect
\citep[see][]{broadhurst95,broadhurst04b}.  In order to make full use
of this method, the unlensed source counts at a given magnitude
threshold must be known accurately. Given the photometric calibration
uncertainties, which at the faint magnitudes one is usually dealing
with easily amount to $0.1\mbox{ mag}$, thus an uncertainty of $\sim
10\%$ in the unlensed source counts is typical. As shown by
\citet{schneider00}, this level of uncertainty removes a great deal of
the power of the magnification method to break the mass-sheet
degeneracy.

 Several attempts have been made recently to measure the cluster mass
  profiles with weak lensing. However, as shown in \citet{clowe01,
  clowe02}, it is extremely difficult to distinguish e.g. isothermal
  from NFW profiles at high significance using weak lensing data
  alone.  The authors also conclude that these difficulties mostly
  arise as a consequence of the mass-sheet degeneracy
  transformation. Therefore additional information needs to be
  included, such as combining the weak lensing data with strong
  lensing (see e.g. \citealp{kneib03,smith04}). Another example was
  given by \citet{sand03} using combined strong lensing and stellar
  kinematics data of the dominating central galaxy. This approach
  offers valuable extra constraints, however a detailed strong-lens
  modelling is required \citep{bartelmann04,dalal03b}.

In this paper we use a combined strong and weak lensing mass
reconstruction to determine the mass and the mass distribution of
clusters. We reconstruct the gravitational potential $\psi$, since it
locally determines both the lensing distortion (for weak lensing) as
well as the deflection (for strong lensing). The method extends the
idea from \citet{bartelmann96} and \citet{seitz98}. Its novel feature
is that we directly include strong lensing information. Further, weak
lensing reconstruction is extended to the critical parts of the
cluster and we include individual redshift information of background
sources as well as of the source(s) being multiply imaged. This allows
us to break the mass-sheet degeneracy and accurately measure the
cluster mass and mass distribution. The method is tested using
simulations, and in \citet{bradac04b} (hereafter
\citetalias{bradac04b}) we apply it to 
the cluster {\RXJ}.  In this paper we first briefly present
the basics of gravitational lensing in Sect.~\ref{sc:gravl}. In
Sect.~\ref{sc:cm} we give an outline of the reconstruction method
(detailed calculations are given in the Appendix). We test the method 
using N-body simulations, and we present the results in
Sect.~\ref{sc:nbody}. The conclusions and summary are the subject of 
Sect.~\ref{sc:concl}.

\section{Gravitational lensing preliminaries}
\label{sc:gravl}
Throughout this paper we follow the notation of
\citet{bartelmann00}, who give a detailed account of
the gravitational lensing theory presented here.

We start by considering a lens having a projected surface mass
density $\Sigma(\vec \theta)$, where $\vec \theta$ denotes the
(angular) position in the lens plane. We define the
dimensionless surface density $\kappa(\vec \theta)$ for a fiducial
source located at a redshift $z \to \infty$ and a lens (deflector) 
at $z=z_{\rm d}$
\begin{equation}
\label{eq:1}
\kappa(\vec \theta) = \frac{\Sigma(\vec \theta)}{\Sigma_{\rm cr}}
\;\mbox{ where } \; \Sigma_{\rm cr} = \frac{c^2}{4\,\pi\,G}\frac{D_\mathrm{\infty}}{D_\mathrm{d} D_\mathrm{d,\infty}} \; .
\end{equation}
$D_{\infty}$, $D_\mathrm{d}$, and
$D_\mathrm{d,\infty}$ are the angular diameter
distances between the observer and the source at $z\to\infty$, 
the observer and the
lens, and the lens and the source, respectively. The choice of scaling 
with $\Sigma_{\rm cr}$ is motivated by the fact
that lenses with $\Sigma \ge \Sigma_{\rm cr}$ (i.e. $\kappa \ge 1$) 
are strong enough to form multiple
images. In this paper, however,  
{\it strong} lensing refers to multiple imaging only,
while {\it weak} lensing means that the lensing effect is
treated as statistical in nature, although it is also applied to the
lens region with $\kappa > 1$, traditionally called the strong lensing
regime.

We define the deflection potential $\psi(\vec \theta)$ 
\begin{equation}
\label{eq:2}
\psi(\vec\theta) = \frac{1}{\pi}\int_{\Re{}^2}{\rm d}^2\theta'\,
  \kappa(\vec\theta')\,\ln|\vec\theta-\vec\theta'|\;.
\end{equation}
which is related to $\kappa$ via the Poisson equation 
\begin{equation}
\label{eq:3}
\nabla^2\psi(\vec\theta) = 2\kappa(\vec\theta)
\end{equation}
Also the shear $\gamma = \gamma_1 + {\rm i} \gamma_2$ and the deflection
angle $\vec \alpha$ are related to $\psi$, where 
\begin{equation}
  \gamma_1 = \frac{1}{2}(\psi_{,11}-\psi_{,22})\;,\quad
  \gamma_2 = \psi_{,12}\;,\quad \vec \alpha = \nabla \psi\;.
\label{eq:6}
\end{equation}
The relations \eqref{eq:3} and \eqref{eq:6} are written for the source
at redshift $z\to \infty$; we note, however, that they hold for any redshift.
Since we will work with sources at different redshifts (both for strong
as well as for weak lensing), we factorise
the redshift dependence of the lens convergence $\kappa$, the shear
$\gamma$, and the deflection angle $\vec \alpha$ by
\begin{align}
 \nonumber
  \kappa(\vec\theta, z) & {} = Z(z) \kappa(\vec\theta) \; , &
  \gamma(\vec\theta, z) & {} = Z(z) \gamma(\vec\theta) \; , &\\
   \label{eq:9}
  \vec \alpha(\vec\theta, z)  & {} =  Z(z) \vec \alpha(\vec\theta)\; .& 
\end{align}
$Z(z)$ is the so-called ``cosmological weight'' function:
\begin{equation}
  \label{eq:10}
  Z(z) \equiv \frac{D_{\infty} D_{\mathrm{d,s}}}{D_{\mathrm{d}, \infty} D_{\mathrm{s}}} \, \mathrm{H}(z-z_\mathrm{d}) \; ,
\end{equation}
where $D_{\mathrm{d,s}}$ and $D_\mathrm{s}$, are the angular diameter
distances between the lens and the source, and the observer
and the source at a redshift $z$, respectively. 
$\mathrm{H}(z-z_\mathrm{d})$ is the Heaviside step function and
accounts for the fact that galaxies located at $z<z_{\rm d}$ are not lensed.\footnote{To evaluate the angular diameter distances we assume
 the $\Lambda$CDM cosmology with $\Omega_{\rm m} = 0.3$, $\Omega_{\Lambda} =
0.7$, and Hubble
constant $H_0 = 70 {\rm km \:
  s^{-1}\:Mpc^{-1}}$.}

In the case of weak lensing, the information on the lens potential is
contained in the transformation between the source ellipticity
$\epsilon^{(s)}$ and image ellipticity $\epsilon$. It is given as a
function of reduced shear $g(\vec \theta, z)$ \citep[see][]{seitz97}
\begin{equation}
  \label{eq:7}
  \epsilon^{(s)} =
  \begin{cases}
    \dfrac{\epsilon - g(\vec \theta, z)}{1 - g^*(\vec \theta,
      z)\epsilon} &
\text{for $\bigl\lvert
      g(\vec\theta, z) \bigr\rvert \leq 1 \; ,$} \\[1.5em]
    \dfrac{1 - g(\vec \theta, z)\epsilon^*}{\epsilon^* - g^*(\vec \theta, z)} &
\text{for $\bigl\lvert
      g(\vec\theta, z) \bigr\rvert > 1 \; .$}
  \end{cases}
\end{equation}
where 
\begin{equation}
g(\vec \theta, z) = \frac{Z(z)\gamma(\vec \theta)}{1-Z(z)\kappa(\vec
  \theta)}\; .
  \label{eq:8}
\end{equation}

Galaxies are intrinsically elliptical, and therefore one cannot disentangle
the effect of lensing from the intrinsic properties 
in \eqref{eq:7} using a single galaxy
image. Hence, the weak lensing effect needs to be treated in
statistical sense. More precisely, we can Taylor expand the expression  
\eqref{eq:7} (e.g. for the case of  $\abs{g(\vec\theta, z)} \le 1$)
and recall that $\abs{\epsilon^{(s)}} < 1$ by definition to obtain
\begin{eqnarray}
  \label{eq:11}
\nonumber
\epsilon(z) &=& \frac{\epsilon^{(s)}+g(\vec \theta, z)}{1+g^*(\vec \theta, z)
\epsilon^{(s)}} \\
&=& \rund{\epsilon^{(s)}+g(\vec \theta,
z)}\sum_{k=0}^{\infty}(-1)^k (g^*(\vec \theta, z))^k (\epsilon^{(s)})^k
\; .
\end{eqnarray}
A similar expansion can be obtained for the case 
$\abs{g(\vec\theta, z)} > 1$.
If we assume
that the intrinsic ellipticity distribution has moments $\bigl\langle {\epsilon^{(s)}}^k \bigr\rangle = 0$ for each
$k$ except $k=0$ we get the known expression for 
the expectation
value of the image ellipticity at redshift $z$  
\begin{equation}
  \label{eq:12}
  \bigl\langle \epsilon(z) \bigr\rangle = \begin{cases}
    g(\vec\theta, z) & \text{if $\bigl\lvert g(\vec\theta, z) \bigr\rvert < 1 \
,$} \\[1em]
    \frac{\displaystyle 1}{\displaystyle g^*(\vec\theta, z)}
    & \text{otherwise$\; .$}
  \end{cases}
\end{equation}
This relation is particularly simple due to the convenient definition
of $\epsilon$. In the approximation $\kappa \ll
  1$, $\abs{\gamma} \ll 1$  (thus $\abs{g} \ll 1$) the
  expectation value is given by
$\bigl\langle \epsilon(z) \bigr\rangle = \gamma(\vec\theta,z)$.

\section{The cluster mass reconstruction method}
\label{sc:cm}

The idea of combining strong and weak lensing constraints is not new,
it has been previously discussed by \citet{abdelsalam98},
\citet{kneib03}, \citet{smith04}, and others. The method presented
here, however, has some important differences. For example in
\citet{abdelsalam98} the authors reconstruct the pixelized version of
the surface mass density $\kappa$. A similar method for strong lensing
constraints only has recently also been presented by \citet{diego04a}.
We argue, however, that using the potential $\psi$ is favourable,
since $\kappa$, $\gamma$, and $\vec\alpha$ locally depend upon the
potential $\psi$ -- c.f. \eqref{eq:3}, \eqref{eq:6} -- and all can be
quantified from the latter.  $\gamma$ and $\vec \alpha$, on the other
hand, are non-local quantities of $\kappa$. In other words, the mass
density on a finite field does not describe the shear and the
deflection angle in this field. If a finite field is used, one usually
employs Fourier analysis; in this case, $\gamma$ in fact corresponds
to original $\kappa$ plus all its periodic continuations.

Further, even though not easy to implement, we decided
to keep the parametrisation of the mass-distribution as general as
possible. In \citet{kneib03} and \citet{smith04}, on the other hand, 
the strong and weak lensing constraints were compared in
a Bayesian approach in the form of simple, parametrised models. In
addition, the weak lensing constraints were not used
to the very centre of the cluster and redshifts of individual sources
were not included.

\subsection{The outline of the method}
\label{sc:cmoutline} 

The main idea behind the method is to describe the
cluster mass-distribution by a fully general lens, 
using the values of the deflection potential
$\psi$ on a regular grid. We then define a  penalty function $\chi^2$ and
minimise it with respect to the values of $\psi_k$.  
The convergence $\kappa$, the shear $\gamma$, and the deflection angle 
$\vec\alpha$ at an arbitrary position in the field are 
obtained by finite differencing and bilinear interpolation
methods. The number of grid points we use for $\psi_k$ is $(\ngx+2) \times
(\ngy+2)$; the extension  by one row and one column at
each side is needed to perform the finite differencing 
at each inner $\ngx \times \ngy$ grid point.

We define the
$\chi^2$-function as follows
\begin{equation}
  \label{eq:13a}
  \chi^2(\psi_k) =    \chi_{\epsilon}^2(\psi_k)  + \chi_{\rm M}^2(\psi_k) + \eta R(\psi_k) \; .
\end{equation}
$\chi_{\epsilon}^2(\psi_k) $ contains information from statistical weak
lensing, whereas in $\chi_{\rm M}^2(\psi_k)$ we include the multiple imaging
properties. $R(\psi_k)$ is a regularisation term multiplied by the
regularisation parameter $\eta$. The regularisation 
is a function of the potential and disfavours small-scale fluctuations
in the surface mass density.

In order to find the minimum $\chi^2$ solution, we solve
\begin{equation}
  \label{eq:13aa}
 \frac{\partial \chi^2(\psi_k)}{\partial \psi_k} = 0\; .
\end{equation}
This is in general a non-linear set of equations, which we solve in
an iterative manner. We linearise this system and in the first step we
start from some
trial solution, to calculate its non-linear terms 
(see the Appendix for details). 
We solve the corresponding equations and repeat this 
procedure until a convergence is achieved.
Inverting the resulting matrix of $\sim \rund{\ngx \times \ngy}^2$ 
elements for finding a solution of the linear system is difficult in
general even for
grids with a small number of cells. However, as it turns out, the
resulting matrix is sparse and the system is therefore computationally 
inexpensive to solve.

\begin{figure*}[ht]
\centering
\includegraphics[width=0.6\textwidth]{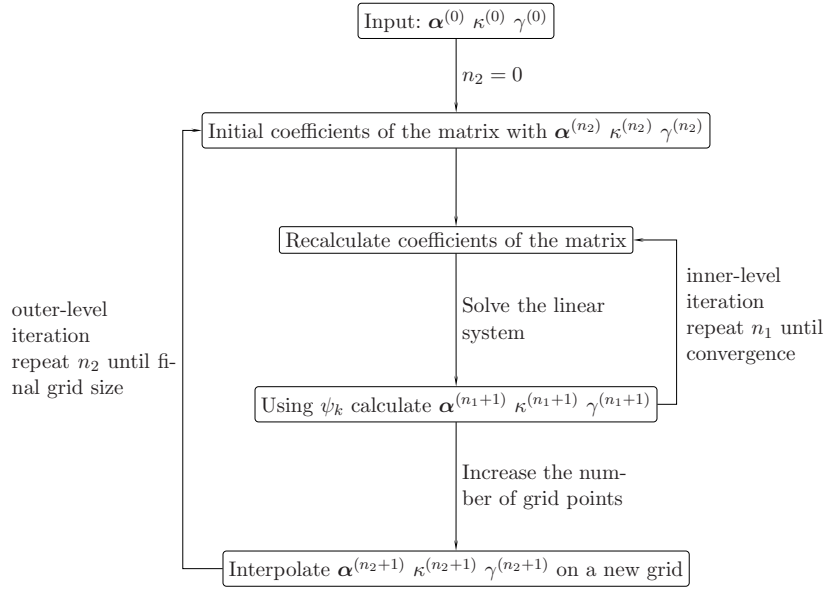}
\caption{The outline of the two-level iteration process.}
\label{fig:method}
\end{figure*}

The reconstruction is performed in a two-level iteration process,
outlined in Fig.~\ref{fig:method}. We will refer to 
the iteration process mentioned above for solving the linear system of 
equations as inner-level, where steps $n_1$ are repeated until
convergence of $\kappa$. 
The outer-level iteration is performed 
for the purpose of regularisation (as described in detail
in Sect.~\ref{sc:chi2}).  In order to penalise small-scale
fluctuations in the surface mass density, we
start the reconstruction with a coarse grid (large cell size). 
Then for each $n_2$ step
we increase the number of grid points in the field  
and compare the new reconstructed
$\kappa^{(n_2)}$ with the one from the previous iteration
$\kappa^{(n_2-1)}$ (or with the initial input value $\kappa^{(0)}$ for
$n_2=0$). The
second-level iterations are performed until the final grid size is
reached and convergence is achieved.   

\subsection{Technical aspects}
\label{sc:technical}
In this section we will briefly describe some technical aspects of how
we calculate the lensing quantities $\kappa$, $\gamma$, and
$\vec\alpha$ at an arbitrary position within the field 
from the potential $\psi_k$ on a grid.

We use the finite differencing method with 
9 grid points to calculate $\kappa$, 5 points for $\gamma$, and 4
points for $\vec \alpha$ \citep[see][]{abramowitz}. 
The coefficients used for $\kappa$ and $\gamma$ are given in
Fig.~\ref{fig:coeff}, the case of $\vec \alpha$ is discussed in 
Appendix~\ref{sc:appendixstrong}. To
evaluate $\kappa(\vec \theta)$, $\gamma(\vec \theta)$, and $\vec
\alpha(\vec \theta)$ at a position $\vec\theta$ within the field, bilinear
interpolation is used.

Note, that the dimensionality of the problem is not $(\ngx+2) \times
(\ngy+2)$. Because the transformation $\psi(\vec \theta) \to \psi(\vec
\theta) + \psi_0 + \vec a \cdot \vec \theta$ leaves $\kappa$ and
$\gamma$ invariant, the potential needs to be fixed at three points
\citep[see][]{seitz98,bartelmann96}. These thus fix the constant and
linear term in the invariance transformation.  If this is not the
case, a minimum in $\chi^2$ would correspond to a three-dimensional
subspace of possible solutions.  The choice of the three points, and
the corresponding values of the potential are arbitrary. Although the
transformation $\psi(\vec \theta) \to \psi(\vec \theta) + \vec a \cdot
\vec \theta$ changes the deflection angle $\vec \alpha$, it only
causes a translation of the source plane, which is not an
observable. Therefore, even in the presence of strong lensing, three
points of the potential need to be held fixed.

The mass-sheet degeneracy transformation of the potential is given
by $\psi \to \psi' = (1-\lambda)
\vec\theta^2 / 2+ \lambda\psi$. However since we aim at lifting this
degeneracy, in contrast to
\citet{seitz98} the potential $\psi_k$, is {\it not} held fixed at an
additional, fourth point. 
The dimensionality of the problem is thus 
$N_{\rm dim} = (\ngx+2)(\ngy+2) - 3$.

\begin{figure*}[ht]
\begin{minipage}{17cm}
\begin{center}
\includegraphics[width=0.9\textwidth]{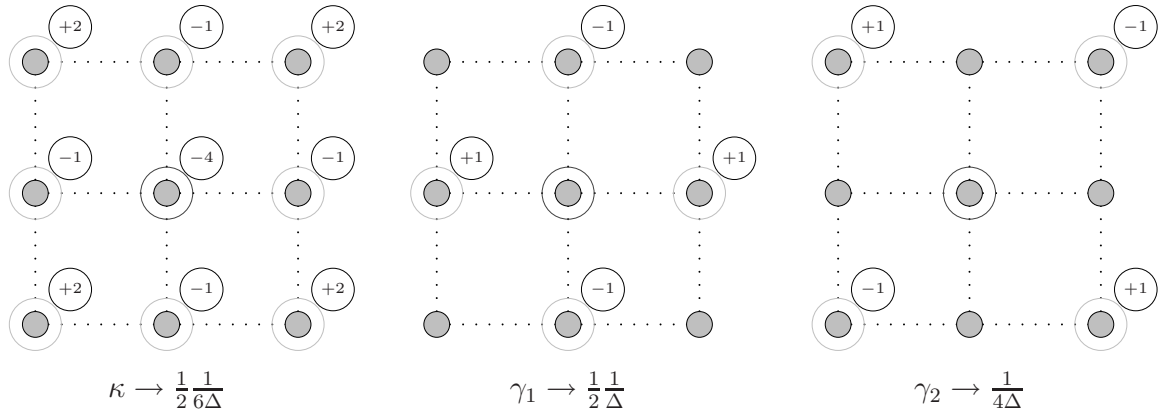}
\end{center}
\end{minipage}
\caption{The finite differencing coefficients for $\kappa$ (left),
  $\gamma_1$ (middle) and $\gamma_2$ (right). E.g. for $\kappa$ we use
  a formula including 9 points, the multiplicative factor is given at
  the bottom, the individual coefficients in the circle. Thus for the
  middle point $(0,0)$ we get $\kappa(0,0) =
  \frac{1}{2}\frac{1}{6\Delta} \big[ 2\eck{\psi(-1,1) +
  \psi(1,1) + \psi(-1,-1) + \psi(1,-1)} -$
  $\eck{\psi(0,1) + \psi(-1,0) + \psi(1,0) + \psi(0,-1)} - 4
  \psi(0,0)\big]$. }
\label{fig:coeff}
\end{figure*}

\subsection{The $\chi^2$-function}
\label{sc:chi2}

In this section we will describe contributions to the
$\chi^2$-function, starting with the statistical weak lensing.

For $N_\mathrm{g}$ galaxies with measured
ellipticities $\epsilon_i$ we define the $\chi_{\epsilon}^2$ as 
\begin{equation}
  \label{eq:13b}
  \chi_{\epsilon}^2(\psi_k) = \sum_{i=1}^{N_\mathrm{g}} \frac{\abs{\epsilon_i -
      \langle \epsilon \rangle}^2}{\sigma_i^2} \; ,
\end{equation}
where
\begin{equation}
  \label{eq:14}
  \sigma_i^2 = \rund{1- \bigl\lvert \langle \epsilon \rangle
  \bigr\rvert^2}^2 \sigma^2_{\epsilon^{\rm s}}
  + \sigma^2_{\rm err}\; .
\end{equation}
Note that $\langle \epsilon \rangle$ refers to the expectation value
  of the ellipticity over redshift space, not to an ensemble average
  over different galaxies and is derived from \eqref{eq:12}.

In \citet{bradac03b}\footnote{We would like to point out here that
  there is a typo in that paper. All $w_i=1/\sigma_i$ in the text and
  plot labels need to be replaced by $w_i = 1/\sigma_i^2$, which has
  been used in the calculations.}  we argue that $\chi_{\epsilon}^2$
can give biased results for lenses for which many galaxies have
$\abs{g} \simeq 1$. It would be better to work with a log-likelihood
function with a probability distribution that properly describes the
distribution of observed ellipticities.  Unfortunately, such an
approach is not viable here (as will become obvious later on).
However, in general clusters do not have a large fraction of galaxies
with $\abs{g} \simeq 1$ and we show in \citet{bradac03b} that for
these lenses the $\chi^2$-minimisation is sufficient.

One of the major strengths of this statistical weak lensing reconstruction
technique is the possibility to simultaneously include
constraints from multiple image systems to the weak lensing data in a
relatively straightforward manner. The simplest approach to strong 
lensing is to perform the so-called 
``source plane'' modelling; i.e. to
minimise the projected source position difference. 
Consider a multiple image system with the source at redshift
$z_{\rm s}$ and with $N_{\rm M}$ images located
at $\vec \theta_{m}$. We define $\vec b_{m} = \vec \theta_{m} - Z(z_{\rm s}) \vec
\alpha(\vec \theta_m) - \vec \beta_{\rm s}$ and the corresponding $\chi^2$-function is given by
\begin{equation}
  \label{eq:17b}
\chi_{\rm M}^2 = \sum_{m=1}^{N_{\rm M}} \vec b_{m}^{\rm T}
\mathcal{S}^{-1} \vec b_{m}\; ,
\end{equation}
where $\mathcal{S}$ is the covariance matrix. $\vec \beta_{\rm s}$ is
the average source position and for simplicity we calculate it using
the deflection angle information from the previous iteration
$n_1-1$. If the measurement errors on image positions are distributed
isotropically, $\mathcal{S}$ reduces to a diagonal matrix given by
$\mathcal{S} = {\rm diag}(\sigma_{\rm s,1}^2,\sigma_{\rm s,2}^2)$ with
$\sigma_{\rm s,1}^2=\sigma_{\rm s,2}^2$. $\sigma_{{\rm s},1}$ and
$\sigma_{{\rm s},2}$ are the errors on image positions, projected onto
the source plane. For simplicity, however, we do not perform a
projection of the error ellipse from the image plane onto the source
plane. Instead, we keep $\mathcal{S}$ constant throughout the
reconstruction.  Therefore we avoided the numerical problem of a
diverging $\chi_{\rm M}^2$ function if one of the images lies at the
position of the critical curve for the corresponding redshift.

We are aware of the fact that the approach we use is not optimal (see
e.g. \citealp{kochanek04_sf}).  If only multiple imaging is used, the
resulting best-fit model is biased towards high magnification factors,
since errors on the source plane are magnified when projected back to
the image plane (this information we do not use).  In our case,
however, the model also needs to take into account the constraints
from statistical (weak) lensing and therefore the high magnification
models are in fact discarded. In addition, if e.g. one considers the
errors matrix in the image plane to be diagonal, the corresponding
matrix in the source plane would have large off-diagonal terms.
Throughout this paper we therefore consider the errors in the source
plane to be isotropic, since this may in fact be a better
approximation, as sources are on average more circular than their
lensed images. In practice the location of the multiple images are
usually known very accurately, leading to a very narrow minimum of
$\chi_{\rm M}^2$ in the parameter space. In practice, multiple image
constraints are satisfied nearly perfectly and exact values of errors
on image positions are of lesser importance.

\begin{figure}[ht]
\begin{center}
\includegraphics[width=0.25\textwidth]{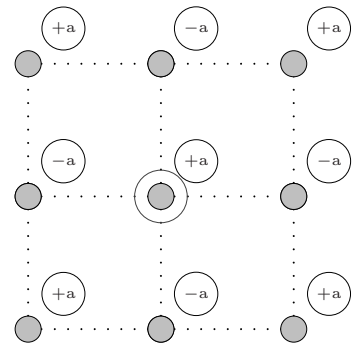}
\end{center}
\caption{The intrinsic invariance of the method. If we alternately
  add/subtract a constant $a$ along the diagonals the shear $\gamma$ does
  not change (cf. Fig.~\ref{fig:coeff}), but $\kappa$ changes in the
  similar way with a constant now being  $2a/3$ .}
\label{fig:coeff2}
\end{figure}

Since the minimisation of $\chi_{\epsilon}^2$ can lead to a potential
that reconstructs the noise in the data, the solution needs to be
regularised. Even without measurement errors, the intrinsic
ellipticities would still produce pronounced small-scale noise peaks
in the final reconstruction.  In addition, the method presented here
has an intrinsic invariance if no multiple imaging information is used
and the weak lensing approximation $ g \simeq \gamma$ applies. Namely,
we can alternately add/subtract a constant $a$ along diagonals of the
potential (chess-board like structure, as sketched in
Fig.~\ref{fig:coeff2}). This transformation would on the one hand not
affect $\gamma$, but on the other it would cause a similar change
(with a constant $2a/3$) in $\kappa$ -- compare with
Fig.~\ref{fig:coeff}. Thus in the $\abs{g} \ll 1$ regime, where
$\ave{\epsilon} = g \simeq \gamma$ these stripes would show up in the
resulting $\kappa$-map. This problem can, however, be very efficiently
cured with regularisation.

Since we want to measure the cluster mass, the regularisation should
not influence breaking the mass-sheet degeneracy. For example, one of
the possibilities considered by \citet{seitz98} for regularisation
function was $R= \sum_{i,j=1}^{\ngx,\ngy} \abs{\nabla\kappa}^2$.
However, as the authors mentioned, such regularisation would tend to
flatten the profile and therefore affect the mass-sheet degeneracy
breaking. Their maximum-entropy regularisation with moving prior
(i.e. the prior in the regularisation is not kept constant, but
adapted in the process of minimisation) does not flatten the profile,
however it is very cumbersome to express its derivative in linear
terms of $\psi_k$.  Motivated by the success of moving prior in
maximum-entropy regularisation, we choose a very simple prescription
for the regularisation function. We start off by a relatively coarse
grid, since if the number of grid points $N_{\rm dim}$ is much smaller
than the number of galaxies, the resulting reconstruction is not able
to follow the noise pattern.  In each second-level iteration step we
gradually increase the number of grid points and compare the resulting
$\kappa$ map $\kappa^{(n_2)}$ with that from the previous iteration
$\kappa^{(n_2-1)}$ linearly interpolated on a finer grid, thus
\begin{equation}
  \label{eq:20}
R = \sum_{i,j=1}^{\ngx,\ngy}\rund{\kappa_{ij}^{(n_2)} -
  \kappa_{ij}^{(n_2-1)}}^2 \; .
\end{equation}
For the case of $n_2=0$ we use an initial guess for $\kappa$ which can
in practice be obtained from strong lensing, direct finite-field
reconstruction, parametrised model fitting to weak lensing data, or
simply set to $\kappa_{ij}^{(0)}=0$. This form of regularisation is
relatively easy to implement and is in addition very efficient in
removing the stripes (mentioned above) in the final reconstruction. If
enough $n_2$ iteration steps are used this form of regularisation also
does not substantially affect the ability to break the mass-sheet
degeneracy, since the information of initial $\kappa^{(0)}$ is lost
and different initial conditions do not bias the results (if the
signal from lensing is high enough).

Finally a word on the regularisation constant $\eta$. This parameter
should, in theory, be set such to ensure $\chi^2/N_{\rm dof} \sim 1$.
In practice, however, it is difficult to determine its optimal value
(in the critical lensing regime).  As outlined in \citet{geiger98} the
probability distribution of measured ellipticities is not a Gaussian
and therefore the minimum value of $\chi^2$ has no particular
meaning. In practice, setting $\eta$ such that the resulting
$\chi^2/N_{\rm dof} \sim 1$ (where $N_{\rm dof}$ is in our case the
number of galaxies used for strong and weak lensing) is valid is a
good guess for this parameter. In addition, one adjusts $\eta$ low
enough for the method to have enough freedom to adapt to the
information in the data and large enough for not allowing the
solutions that follow the noise pattern. As a rule of thumb it is
usually better to set $\eta$ high and increase the number of
iterations and hence allowing $\kappa$ to change only slowly. Since
the reconstruction is done in a two-level iteration and in addition
multiple-image information is included, the method can successfully
adapt to the data and the results are not very sensitive to the
precise value of $\eta$. The resulting smoothness level of the mass
maps should reflect the quality of data. The ``smoothing scale''
depends upon the combination of the grid size and regularisation. The
final potential map should be void of any structures on scales smaller
than the mean separation between galaxies used for weak lensing. We
will shortly return to this point in Sect.~\ref{sc:res1}.

\subsection{Initial conditions}
\label{sc:ini}
For the purpose of the regularisation, given by  \eqref{eq:20}, we
need to employ the initial conditions. In addition, if a realistic model
is used for the initial conditions (and as trial solution), it is 
very helpful to break the internal degeneracy, i.e. to distinguish
galaxies that have $\abs{g} \le 1$ from those with $\abs{g} > 1$ in
the first step, and thus allow for a faster convergence. Breaking
this degeneracy is desirable, 
since otherwise the method has difficulties in ``climbing''
over the $\abs{g} = 1$ region (especially if {\it no} multiple-image
 systems are included). Since we use individual redshifts of
background galaxies we do not have well-defined critical curves
 (i.e. positions in the source plane where  $\abs{g} = 1$), 
as their position depends on the source redshift. In spite of this fact, 
the transition still poses a difficulty.

In our case different initial conditions are employed. For the initial
model $\kappa^{(0)}$ we use three different scenarios: $\kappa^{(0)} =
0$ (and $\vec\alpha^{(0)} = \vec 0$, $\gamma^{(0)} = 0$) 
across the whole field (hereafter {\it
  I0}), $\kappa^{(0)}$ taken from the best fit non-singular isothermal
ellipsoid model NIE for the multiple image system described in
Sect.~\ref{sc:mult} (hereafter {\it IM}) and a non-singular isothermal
sphere model NIS with scaling and core radius being the same as in
{\it IM} (hereafter {\it IC}). The same models are used also to obtain
the initial coefficients of the linear system (see Appendix) (for {\it
  I0} we use $\gamma_{1,2}=0$). These different initial models help us
to explore the effects of regularisation and the capability of the
reconstruction method to adapt to the data.

\section{Cluster mass reconstruction from simulated data}
\label{sc:nbody}
\subsection{Mock catalogues}
\label{sc:mock}
\begin{figure*}[ht]
\begin{minipage}{17cm}
\begin{center}
\includegraphics[width=0.9\textwidth]{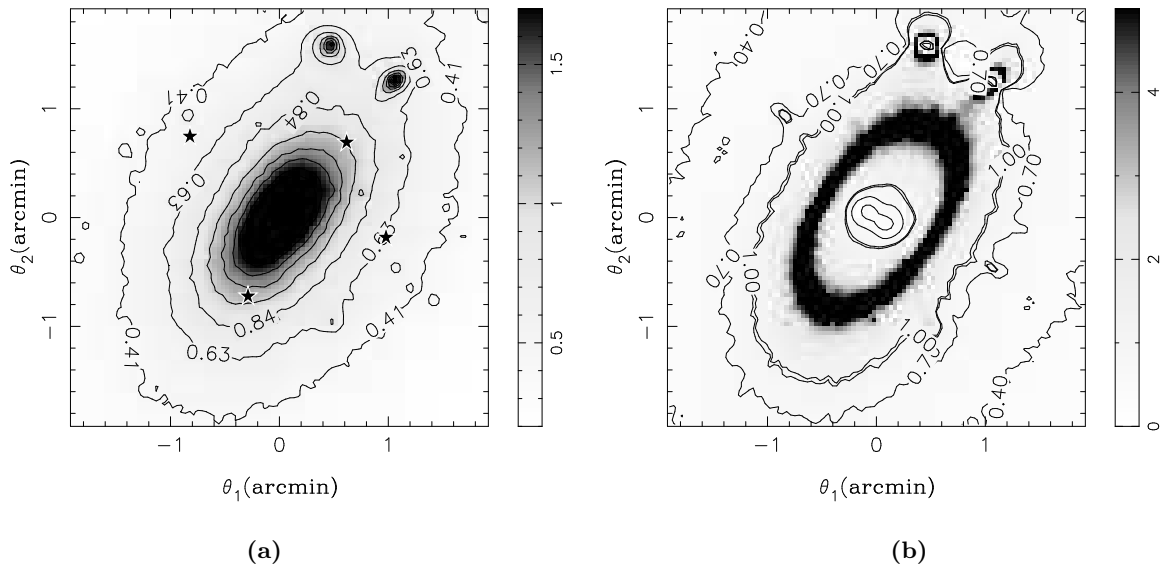}
\end{center}
\begin{minipage}{8.5cm}
\centerline{\bf (a)}
\end{minipage}
\begin{minipage}{8.5cm}
\centerline{\bf (b)}
\end{minipage}
\end{minipage}
\caption{The gravitational lensing properties of a simulated cluster
  used for generating mock catalogues for statistical weak lensing and
  for the multiple image system. {\bf a)} The surface mass density $\kappa$,
  {\bf b)} the absolute value of the reduced shear $\abs{g}$, both 
for a source at
  $z_{\rm s} \to \infty$, are given in gray-scale and contours. The stars
  in {\bf a)} denote the image positions of a four-image system at
  $z_{\rm s} = 1.76$ which we use
  for the reconstruction. }
\label{fig:simulatedlens}
\end{figure*}

We generate mock catalogues using a cluster from the
high-resolution N-body simulation by \citet{springel01}. The cluster
is taken from the S4 simulation (for details see the aforementioned
paper) and was simulated in the framework of the $\Lambda$CDM
cosmology with density parameters $\Omega_{\rm m} = 0.3$ and
$\Omega_{\Lambda} = 0.7$, shape parameter $\Gamma = 0.21$, the
normalisation of the power spectrum $\sigma_8=0.9$, and Hubble
constant $H_0 = 70 \;{\rm km \: s^{-1}\:Mpc^{-1}}$. The cluster
simulation consists of almost 20 million particles, each with a mass
of $4.68\times10^7 \; M_{\odot}$ and a gravitational softening length
of $0.7\:h^{-1}\:{\rm kpc}$. Due to the high mass resolution, the
surface mass density $\kappa$-map can be obtained by directly
projecting the particles (in our case along the $z$-axis) onto a
$1024^2$ grid (of a side length $4 \mbox{ Mpc}$) using the NGP
(nearest gridpoint) assignment.

In what follows we try to generate the weak and strong lensing data to
resemble as close as possible the data on the cluster {\RXJ} we will
use in \citetalias{bradac04b}.  The surface mass density of the
cluster is therefore scaled to have a sizeable region where
multiple imaging is possible within a $3.8 \times 3.8 \mbox{
arcmin}^2$ field, for sources at redshifts $z\gtrsim 1$ and a cluster
at $z_{\rm d} = 0.4$. The Einstein radius for a fiducial source at
$z\to\infty$ is roughly $\theta_{\rm E} \sim 1^{\prime}$, giving a
line-of-sight integrated mass within this radius of $\sim
5\times10^{14} M_{\odot}$. The cut-out of the resulting $\kappa$ map
(for $z\to\infty$, thus $Z(z) = 1$) we use can be seen in
Fig.~\ref{fig:simulatedlens}a.

The lensing properties are calculated as described in detail in
\citet{bradac03}. The Poisson equation for the lens potential $\psi$
-- c.f. Eq.~\eqref{eq:3} -- is solved on the grid in Fourier space
with a DFT (Discrete Fourier Transformation) method using the FFTW
library written by \citet{frigo98}. The two components of the shear
$\gamma_{1,2}$ and the deflection angle $\vec \alpha$ are obtained by
finite differencing methods applied to the potential $\psi$. These
data are then used to generate the weak lensing catalogues as well as
the multiple image systems.  The absolute value of the reduced shear
(again for a source with $z\to\infty$) is shown in
Fig.~\ref{fig:simulatedlens}b.

The weak lensing data are obtained by placing $N_{\rm g}$ galaxies 
on a $3.8 \times 3.8 \mbox{ arcmin}^2$ field.  
We have simulated two different catalogues, one
with $N_{\rm g} = 148$ galaxies with positions corresponding to those
from R-band weak lensing data of the cluster {\RXJ} and
one with  $N_{\rm g} = 210$ galaxies corresponding to the I-band
data used in \citetalias{bradac04b}. 
In this way we simulate the effects
of ``holes'', resulting from cluster obscuration and bright stars in
the field. 

The
intrinsic ellipticities $\epsilon_{\rm s}$ are drawn from a Gaussian
distribution, each component is characterised by  $\sigma = \sigma_{\epsilon^\mathrm{s}}
= 0.2$. We use the same redshifts as those measured in the R and
I-band data, respectively. The catalogues have average redshifts for
background sources of $\ave{z_{\rm I}} = 1.18$ and $\ave{z_{\rm R}} =
1.14$. The corresponding cosmological weights are evaluated assuming the
$\Lambda$CDM cosmology (the same parameters are used as for the cluster
simulations). 

The measurement errors $\epsilon^{\rm err}$ on the observed
ellipticities are drawn from a Gaussian distribution with $\sigma
=\sigma_\mathrm{err} = 0.1$ (each component) and added to the lensed
ellipticities. We considered also the measurement errors on the
redshifts of the galaxies to simulate the use of photometric
redshifts.  These have $\sigma_\mathrm{zerr} = 0.1 \rund{1+z_i}$
\citep[see][]{bolzonella00}; in adding the errors we ensured that the
resulting redshifts are always positive. We have also simulated the
presence of outliers in the redshift distribution, $10\%$ of our
background sources (chosen at random) are considered outliers, 
for these we randomly
add/subtract $\Delta z = 1$ to their redshifts (which already include
random errors). The lensed ellipticities are obtained using
\eqref{eq:11} and interpolating the quantities $\kappa$, and $\gamma$
at the galaxy position using bilinear
interpolation, considering the redshifts including errors. 
In contrast, for the purpose of reconstruction we then consider
galaxies to be at their "original" redshifts (thus equal to the
observed redshifts in the data of {\RXJ}).

\subsection{Multiple imaging}
\label{sc:mult}
To obtain a four-image system from the simulation 
we use the method described in detail in \citet{bradac03}. 
With the {\tt MNEWT} routine from \citet{numrec_c} we solve 
the lens equation for a given source position inside the
asteroid caustic. The source is assumed at a redshift of 
$z_{\rm s} = 1.76$. Once we have the image positions, their magnifications are
calculated and the fifth image is eliminated, since it is usually too
dim and would not be observable.

The errors on image positions can be conservatively
estimated (for the data we use in \citetalias{bradac04b}) 
to $ \sim 0.\arcsecf3$. Since we need
errors in the source plane, we set them by a factor of five smaller
(in agreement with the average magnification factor for this system), i.e.
$\sigma_{{\rm s},m} = 0.\arcsecf06$ for both coordinates (see discussion in Sect.~\ref{sc:chi2}).

We also use this system to obtain one of the $\kappa^{(0)}$ models,
needed for the purpose of strong and weak lensing reconstruction. We
perform image plane minimisation and fit an NIE 
model \citep{ko94a}, given by 
\begin{equation}
  \kappa(\vec \theta') = \frac{b_0}{2\sqrt{\frac{1+\abs{\epsilon_{\rm
            g}}}{1-\abs{\epsilon_{\rm g}}} \rund{r_{\rm c,nis}^2 +
        (\theta'_1)^2} + (\theta'_2)^2}} \; ,
\label{eq:nis}
\end{equation}
where $\theta'$ is calculated w.r.t. the semi-major axis 
of the cluster surface mass density. We allow the scaling $b_0$,
ellipticity $\abs{\epsilon_{\rm g}}$ and position angle $\phi_g$ to
vary.  The best fit model for this system has values of $\{b_0,
\abs{\epsilon_{\rm g}}, \phi_g\} = \{0.\arcminf97,0.30,1.01\}$. We fix
the core radius $r_{\rm c}$ to $0.\arcminf1$. For model fitting
we use \texttt{C-minuit} \citep{james75}, a
routine which is part of the CERN Program Library.
\begin{figure*}[tbp]
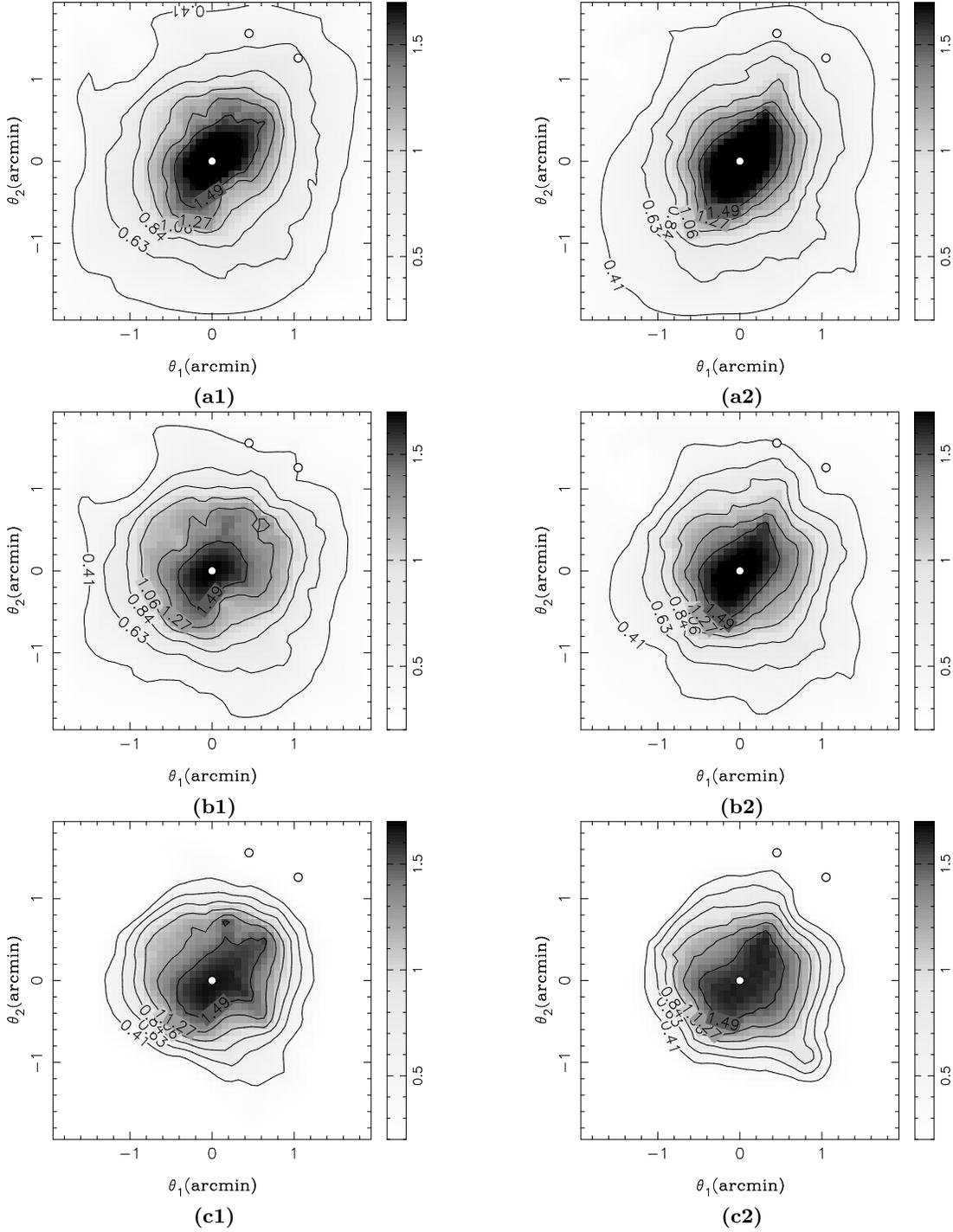

\begin{minipage}{17cm}
\begin{minipage}{8cm}
\begin{center}
\includegraphics[width=0.8\textwidth]{2233fig5a.ps}
\centerline{{\bf (a1)}}
\includegraphics[width=0.8\textwidth]{2233fig5b.ps}
\centerline{{\bf (b1)}}
\includegraphics[width=0.8\textwidth]{2233fig5c.ps}
\centerline{{\bf (c1)}}
\end{center}
\end{minipage}
\begin{minipage}{8cm}
\begin{center}
\includegraphics[width=0.8\textwidth]{2233fig5d.ps}
\centerline{{\bf (a2)}}
\includegraphics[width=0.8\textwidth]{2233fig5e.ps}
\centerline{{\bf (b2)}}
\includegraphics[width=0.8\textwidth]{2233fig5f.ps}
\centerline{{\bf (c2)}}
\end{center}
\end{minipage}
\end{minipage}
\caption{$\kappa$-maps obtained from 
combined strong and weak lensing reconstruction of the simulated data. 
 {\bf Left} panels show the reconstructions using $N_{\rm g} = 210$
 galaxies distributed in the same manner as the I-band, while for the
 {\bf right} panels we use  $N_{\rm g} = 148$ galaxies
  distributed in the same manner as the R-band weak lensing data for
  {\RXJ} (see \citetalias{bradac04b}). The galaxies 
  have been lensed by an N-body simulated cluster. Different initial
  conditions are used for the reconstruction. 
In {\bf a1-a2)} we use best fit
  model from the multiple image system  {\it IM} (see
  Sect.~\ref{sc:mult}) in {\bf b1-b2)} we use the  {\it IC} model,
  an NIS model with the same scaling and core radius as  {\it IM}  and in
  {\bf c1-c2)} we use  {\it I0}, i.e. 
  $\kappa^{(0)}=0$ on all grid points.
 The positions of the cluster
centre and two major subclumps are plotted as white circles.}  
\label{fig:nbody_wl}
\end{figure*}

\subsection{Weak lensing mass-reconstruction using simulated data}
\label{sc:res1}  
The mock catalogues are used to test the performance of the
reconstruction method. We obtain the solution of the linear system
with the \texttt{UMFPACK} routine for solving asymmetric sparse
linear systems \citep{umfpack}. We perform 30 second-level iterations,
each time increasing $\ngx$ and $\ngy$ by one, starting with an
initial $20\times20$ grid (since we use regularisation from the first
iteration, we do not need to start the reconstruction with a very
course grid, where the resulting matrix is not sparse and a different
routine should be applied to find a good solution).  The results of
the reconstructions are shown in Fig.~\ref{fig:nbody_wl} for $N_{\rm
  g} = 210$ galaxies distributed in the same manner as the I-band data
and $N_{\rm g} = 148$ galaxies distributed as the R-band data of the
cluster described in \citetalias{bradac04b}.  The initial
regularisation parameter was set to $\eta = 400$ for I-band $\eta =
200$ for R-band data and adapted in each step to ensure $\chi^2/N_{\rm
  dof} \sim 1$. Two different initial values for the regularisation parameter
are used since the numbers of galaxies in the two bands are different.
It is very comforting to observe that the reconstructed maps do not
depend crucially on the initial $\kappa$-model we use.

From the reconstructed maps we estimate the mass within the
$1.\arcminf5$ radius from the centre of the cluster (for a redshift
$z_{\rm d} = 0.4$ this corresponds to $340 {h^{-1}}\mbox{ kpc}$)
projected along the line-of-sight.  For this purpose we generate 10
mock catalogues for each band and did the reconstruction again with
the three different initial models. We list the resulting average mass
obtained from the catalogues in Table~\ref{tab:mass_sim} for both the
I- and R-band mock catalogues.  All the mass estimates are similar;
note, however, that the galaxy catalogues following the I- and R-band
data have galaxies partly in common and the errors are therefore
correlated.  We find the enclosed mass of the simulated cluster to be
$(1.0 \pm 0.1) \times 10^{15} M_{\odot}$, which is very close to the
input value of $M_{\rm s}(<340 {h^{-1}}\mbox{ kpc}) = 0.99 \times
10^{15} M_{\odot}$. The 1-$\sigma$ error is estimated from the
variance of mass determinations from different mock catalogues.

The results show that our method is {\it effectively} able to break
the mass-sheet degeneracy and is, as a consequence, very efficient in
reproducing the cluster mass also at radii significantly larger than
the Einstein radius of the cluster.  It is also very encouraging that
the results are nearly independent of the initial $\kappa^{(0)}$ used
for the regularisation. Note that a single multiple-image system does
not by itself break this degeneracy, we would need at least two
different redshift multiple image systems to break the mass-sheet
degeneracy with strong lensing data alone. In such a case the strong
lensing gives constraints on the mass enclosed within the Einstein
radius for a given source redshift and since the critical curves
depend on the source redshift, we can constrain mass at two different
radii and the degeneracy is broken.  The combiniation of weak and
strong lensing is the more powerful, the more different the redshift
of the source of the multiple images is from the median redshift of
the galaxies from which the weak lensing measurements are obtained,
and the less symmetric the arrangement of multiple images is
w.r.t. the center of the cluster.

Unfortunately we can not resolve both clumps present in the
simulations. This is due to the fact that the number density of
background sources is low and the internal smoothing scale (i.e. the
average distance between two source galaxies) is large; with a number
density of $\sim 100 \mbox{ arcmin}^{-2}$ the
clumps can be easily resolved.

We have also performed additional reconstructions in which we 
multiplied the original values of $\kappa$ of the simulated cluster by
$0.75$ and $1.25$. This enables us to confirm that the agreement between input
mass and reconstructed mass is not just accidental. We
have generated new multiple image systems and new mock catalogues as
before. We do not, however, perform a new strong-lensing reconstruction,
for $\kappa^{(0)}$ we intentionally use the same (i.e. in this case
``wrong'') initial
conditions as before. The old {\it IM} model would not fit the image
positions any longer, since they have changed with the scaling of
$\kappa$. 
The reconstructed masses of the increased $\kappa$ simulation
are in good agreement with the input values. The differences
between different models are comparable (slightly smaller) to the ones 
shown in
Table~\ref{tab:mass_sim}. For the lower $\kappa$ simulation, the
reconstructed values are on average the same as the input value, however
the scatter is larger. This is expected, since the lens in this
case is weaker and the breaking of the mass-sheet degeneracy is
difficult in this case (with the quality of data used here).

As an additional test we also consider a redshift distribution with
$\ave{z} = 1.6$ for the sources used for weak lensing and regenerate
the mock catalogues. The accuracy of the determination of the enclosed
mass increases. However, more importantly, we also better reconstruct
the shape of the mass distribution, since
high-redshift galaxies (when their shape is measured reliably)
contribute most to the signal and improve the accuracy of the
reconstruction.

 \begin{table}[b!]
\caption{Reconstructed cluster mass within a cylinder of $340
{h^{-1}}\mbox{ kpc}$ radius around the cluster centre from simulations
of mock catalogues resembling I-band (left) and R-band (right) weak
lensing data and one 4-image system.  Three different initial
conditions are used.  We use the best fit model from the multiple
image system {\it IM} (see Sect.~\ref{sc:mult}), the {\it IC} model
(NIS with same scaling and core radius as {\it IM}) and {\it I0} with
$\kappa^{(0)}=0$ on all grid points. In the last line the input mass
$M_{\rm s}$ from the simulation is given. The variance of the mass
estimate is given and in brackets we give for
comparison the velocity dispersion of an SIS having the same enclosed
mass within $340 {h^{-1}}\mbox{ kpc}$.}
\label{tab:mass_sim}
\begin{center}
\begin{tabular}{l c c  c c}
\noalign{\smallskip}
\hline
\noalign{\smallskip}
\hline
\noalign{\smallskip}
 & $M_{I,\rm s}$& $[\sigma_{\rm I,SIS}]$ &$M_{R,\rm s}$& $[\sigma_{\rm R,SIS}]$\\
&$[10^{15} M_{\odot}]$& $[{\rm km s^{-1}}]$& $[10^{15} M_{\odot}]$& $[{\rm km s^{-1}}]$\\
\noalign{\smallskip}
\hline
\noalign{\smallskip}
{\it IM} & $1.07 \pm 0.02$&[1720]&$1.04 \pm 0.02$ &[1710]\\
 {\it IC} & $1.04 \pm 0.02$&[1700]&$1.00 \pm 0.02$ &[1670]\\
 {\it I0} & $0.88 \pm 0.03$&[1560]&$0.82 \pm 0.03$ &[1510]\\
\noalign{\smallskip}
\hline
\noalign{\smallskip}
\multicolumn{3}{c}{$M_{\rm s}(<340 {h^{-1}}\mbox{ kpc})$} & \multicolumn{1}{c}{0.99} &[1670]\\
\noalign{\smallskip}
\hline
\end{tabular}
\end{center}
\end{table}

\section{Conclusions}
\label{sc:concl}

In this paper we develop a new method based on
\citet{bartelmann96} to perform a combined weak and strong lensing
cluster mass reconstruction. The particular strength of this method is
that we extend the weak lensing analysis to the critical parts of the
cluster. In turn, this enables us to directly include multiple imaging
information to the reconstruction. Strong and weak lensing
reconstructions are performed on the same scales, in contrast to
similar methods proposed in the past where weak lensing information on
much larger radii than the Einstein radius $\theta_{\rm E}$ was
combined with strong lensing information (see e.g. \citealp{kneib03}).

We test the performance of the method on simulated data and conclude
that if a quadruply imaged system combined with weak lensing data and
individual photometric redshifts is used, the method can very
successfully reconstruct the cluster mass distribution. With a
relatively low number density of background galaxies, $15 \mbox{
  arcmin}^{-2}$, we are effectively able to reproduce the main
properties of the simulated cluster. In addition, with larger number
densities $\sim 100 \mbox{ arcmin}^{-2}$ of background sources,
accessible by HST, the substructures in the cluster can be resolved
and the mass determination further improved.

We determine the enclosed mass within $340 {h^{-1}}\mbox{ kpc}$ of the
simulated cluster to be $(1.0 \pm 0.1) \times 10^{15} M_{\odot}$,
which is very close to the input value of $M_{\rm s}(<340
{h^{-1}}\mbox{ kpc}) = 0.99 \times 10^{15} M_{\odot}$. We have shown,
that with the data quality we use we are effectively able to break the
mass-sheet degeneracy and therefore obtain the mass and
mass-distribution estimates without prior assumptions on the lensing
potential.

In addition, the reconstruction algorithm can be improved in many
ways. First, we use for the multiply imaged system only the
information of the image positions.  The reconstruction method can,
however, be modified to include the morphological information of each
extended source. Instead of using a regular grid, one would have to
use adaptive grids and decrease the cell sizes around each of these
images. This will be a subject of a future work. Second, the
photometric redshift determination does not only give the most likely
redshift given the magnitudes in different filters, but also the
probability distribution for the redshift. This information can be
included in the reconstruction. In addition, source galaxies without
redshift information can be included and different regularisation schemes
can be considered.

Finally, the slight dependence on the initial conditions is getting
weaker the higher the number density of background galaxies and/or
multiple image systems are. In addition, it is of advantage to have a
large spread in the redshift efficiency factors $Z$ of the background
galaxies. For example, deep ACS images of clusters with a usable
number density of $n\sim 120 \mbox{ arcmin}^{-2}$, or future
observations with the James Webb Space Telescope will most likely make
the dependence on the initial conditions negligible.

In \citetalias{bradac04b} we will show the application of this
method on the cluster {\RXJ} and confirm that  a combination of strong and weak
lensing offers a unique tool to pin down the masses of 
galaxy-clusters as well as their mass distributions.

\begin{acknowledgements}
We would like to give special thanks to Volker Springel for providing
us with the cluster simulations.  We would further like to thank
L\'{e}on Koopmans and Oliver Czoske for many useful discussions that
helped to improve the paper.  We also thank our referee for his
constructive comments. This work was supported by the International
Max Planck Research School for Radio and Infrared Astronomy, by the
Bonn International Graduate School, and by the Deutsche
Forschungsgemeinschaft under the project SCHN 342/3--3. MB
acknowledges support from the NSF grant AST-0206286. This project was
partially supported by the Department of Energy contract
DE-AC3-76SF00515 to SLAC.
\end{acknowledgements}

\appendix 
\section{The linear problem for $\psi_k$}
 \label{sc:appendixBij}

In this section we present details of the method outlined in
Sect.~\ref{sc:cm}.
  
We aim to solve the equation 
\begin{equation}
  \label{eq:A1}
  \frac{\partial \chi_{\epsilon}^2(\psi_k)}{\partial\psi_k} +   \frac{\partial
  \chi_{\rm M}^2(\psi_k)}{\partial\psi_k}  +
  \eta \frac{\partial R(\psi_k)}{\partial\psi_k} = 0 \; .
\end{equation}
This is in general a  non-linear system of
equations. We try to solve it in an iterative way by linearising the
equation in terms of $\psi_k$ 
and keeping the non-linear terms fixed at each
iteration step. The resulting system is then written in the form 
\begin{equation}
  \label{eq:17}
\mathcal{B}_{jk} \psi_k = V_j \;,
\end{equation}
where the matrix $\mathcal{B}_{jk}$  and vector $V_j$ contain the
contributions from the non-linear part.  In the following sections we
will describe the contributions to \eqref{eq:A1} in turn.

\subsection{The weak lensing analysis}
 \label{sc:appendixweak}
The $\chi_{\epsilon}^2$ for the weak lensing case is given in
\eqref{eq:13b}. From now on we consider in detail only the
$\abs{g} \le 1$ case; for  $\abs{g} > 1$, the calculations are done in the same
fashion. 
First we plug into  \eqref{eq:13b} the expectation value of observed
ellipticities (i.e. the reduced shear $g$) obtaining
\begin{equation}
  \label{eq:15}
  \chi_{\epsilon}^2(\psi_k) = 
 \sum_{i=1}^{N_\mathrm{g}} \frac{\abs{\epsilon -
  \frac{Z\gamma}{1-Z\kappa}}^2}{\sigma^2}=  \sum_{i=1}^{N_\mathrm{g}}
  \frac{\abs{\epsilon - Z\epsilon\kappa -Z\gamma}^2}{\rund{1-Z\kappa}^2\sigma^2} \; ,
\end{equation} 
where $\kappa$, $\gamma$, and $\sigma$ depend on $\vec \theta_i$ only  and $Z$
depends on the redshift of the  $i$-th source. 
Note that for simplicity we omit
the index $i$ to $\epsilon$, $\kappa$, $\gamma$, $\sigma$ and $Z$,
although these quantities are 
 different for every galaxy. As described in
Sect.~\ref{sc:technical} using finite differencing and bilinear
interpolation, we can write  $\kappa$ and $\gamma$ at each galaxy
position as a linear combination of $\psi_k$. This is expressed in 
the following matrix
notation
\begin{equation}
\gamma_1(\vec \theta_i) = \ga{ik} \psi_k \; , \; \; \; 
\gamma_2(\vec \theta_i) =  \gb{ik} \psi_k \; , \; \; \;
\kappa(\vec \theta_i) = \k{ik} \psi_k \; ,
\label{eq:17a}
\end{equation}
where the matrices $\ga{ik}$,
$\gb{ik}$, and $\k{ik}$ are composed of
numerical factors described in Sect.~\ref{sc:technical}. 
Now we consider the denominator of \eqref{eq:15}
$\hat{\sigma}_{\le}^2 = \rund{1-Z\kappa}^2\sigma^2$ 
fixed at each iteration step 
(the subscript $\le$ denotes the $\abs{g} \le 1$ case) and 
differentiate the following term of \eqref{eq:15}  
 \begin{eqnarray}
\nonumber
&\phantom{+}&\;\frac{1}{2}\frac{\partial}{\partial\psi_k} \frac{\abs{\epsilon -
Z\epsilon\kappa -Z\gamma}^2}{\hat{\sigma}_{\le}^2}  =\\
\nonumber
&\phantom{+}&-\;\frac{Z}{\hat{\sigma}_{\le}^2} \left[\rund{\epsilon_1 -
Z\epsilon_1\kappa -Z
\gamma_1}\rund{\epsilon_1\frac{\partial\kappa}{\partial\psi_k}
+ \frac{\partial\gamma_1}{\partial\psi_k}} \right. \\
\nonumber 
&\phantom{+}&+\;\left. \rund{\epsilon_2 -
Z\epsilon_2\kappa -Z
\gamma_2}\rund{\epsilon_2\frac{\partial\kappa}{\partial\psi_k}
+ \frac{\partial\gamma_2}{\partial\psi_k}}\right] = \\
\nonumber
&\phantom{+}&\phantom{-}\;\;\frac{Z^2}{\hat{\sigma}_{\le}^2} \Big\lbrack \ga{ij}\ga{ik} + \gb{ij} \gb{ik} +
      \epsilon_1 \rund{\ga{ij}\k{ik} + \k{ij}\ga{ik}} \\
\nonumber
&\phantom{+}&+\; \;\epsilon_2
      \rund{\gb{ij}\k{ik} + \k{ij}\gb{ik}} + \rund{\epsilon_1^2 +
      \epsilon_2^2} \k{ij}\k{ik}\Big\rbrack\psi_k \\
&\phantom{+}&- \;\;\frac{Z^2}{\hat{\sigma}_{\le}^2}\eck{ \epsilon_1 \ga{ij} + \epsilon_2
      \gb{ij} +   \rund{\epsilon_1^2 +
      \epsilon_2^2} \k{ij}}
  \label{eq:16}
\end{eqnarray} 
where $\epsilon_1$ and $\epsilon_2$ are the two components of the measured
ellipticity of galaxy $i$ (again omitting the index and we divided
Eq.~\eqref{eq:A1} by two for simplicity). We sum over all
galaxies used for the weak lensing analysis and obtain a 
linear problem for $\psi_k$ at each iteration step. 
The same approach can be used for the $\abs{g}
> 1$ case, where $\hat{\sigma}_{>}^2$ is kept constant and is given by 
$\hat{\sigma}_{>}^2 = Z_i^2 \abs{\gamma}^2 \sigma_i^2$.

\subsection{The strong-lensing term}
 \label{sc:appendixstrong}

Following the prescription from the previous section we now write the
deflection angle in a matrix form 
\begin{equation}
\alpha_1(\vec \theta_m) = \mathcal{D}^{(1)}_{ik} \psi_k \; , \; \; \; 
\alpha_2(\vec \theta_m) =  \mathcal{D}^{(2)}_{ik} \psi_k \; . 
\label{eq:17c}
\end{equation}
Both matrices give the finite differencing form for the gradient of
the potential, in particular we use  the central differencing formula,
i.e. $\alpha_1(0,0) =
\frac{1}{2\Delta} \rund{\psi(1,0) -\psi(-1,0)}$ and   $\alpha_2(0,0) =
\frac{1}{2\Delta} \rund{\psi(0,1) -\psi(0,-1)}$. 

The $\chi^2$ contribution to strong lensing is given in \eqref{eq:17b}.
The source position $\vec \beta_{\rm s}$ is kept constant at every
iteration step, and is evaluated using the deflection angle
information  $\vec \alpha^{(n_1-1)}$ from the previous iteration 
\begin{equation}
\vec \beta_{\rm s} = \frac{1}{N_{\rm M}} \sum_{m=1}^{N_{\rm M}}
\rund{\vec \theta_{\rm m} - Z(z_{\rm s}) \vec \alpha^{(n_1-1)}(\vec
  \theta_m)} \; .
\label{eq:17d}
\end{equation}
We differentiate the following term in $\chi^2_{\rm M}$ (for the $x_1$-coordinate) 
\begin{eqnarray}
\nonumber
&\phantom{=}&\frac{1}{2}
\frac{\partial}{\partial\psi_k} \frac{\rund{\theta_{\rm m,1}- Z(z_{\rm s})
\alpha_1(\vec \theta_m) - \beta_{\rm s,1}}^2}{\sigma_{{\rm s},m(1)}^2}
= \\
&\phantom{=}&-\frac{\rund{\theta_{m,1}  - \beta_{\rm s,1}}\aa{ij} -
      Z(z_{\rm s})\aa{ij} \aa{ik} \psi_k}{\sigma_{{\rm s},m(1)}^2}\; .
\label{eq:17e}
\end{eqnarray}
The expression for the $x_2$-coordinate is obtained by exchanging
$1\rightarrow2$. After summation of both terms over all images $m$ we
get a set of equations which are linear in $\psi_k$ and can be readily
included in \eqref{eq:17}. In principle we could also use $\vec
\alpha(\psi_k)$ in \eqref{eq:17d}, the expression $\partial \chi_{\rm
M}^2 / \partial\psi_k$ would remain linear in $\psi_k$. However since
the first approach is computationally much simpler, we use the former.

\subsection{The final result}
 \label{sc:appendixfinal}

In the previous section we describe how we linearise the
contributions of weak and strong lensing, now we can write the
coefficients in the equation
\eqref{eq:17}. Note that the contribution of the regularisation term (with
$\chi^2$-contribution given in \eqref{eq:20}) is already linear in
$\psi_k$ and therefore the full matrix $\mathcal{B}_{jk}$ 
is  given in the form 
\begin{eqnarray}
\nonumber
\mathcal{B}_{jk} &=& \sum_{i=1}^{N_{\rm gal}}  \left[ a_{11}(i)
\mathcal{G}^{(1)}_{ij} \mathcal{G}^{(1)}_{ik} + a_{22}(i)
\mathcal{G}^{(2)}_{ij} \mathcal{G}^{(2)}_{ik} \right.\\
\nonumber
&\phantom{=}&\qquad+a_{13}(i) \rund{\mathcal{G}^{(1)}_{ij} \mathcal{K}_{ik} +
  \mathcal{G}^{(1)}_{ik} \mathcal{K}_{ij}} \\
\nonumber
&\phantom{=}&\qquad+ a_{23}(i)\rund{\mathcal{G}^{(2)}_{ij} \mathcal{K}_{ik} +
  \mathcal{G}^{(2)}_{ik} \mathcal{K}_{ij}}\\
\nonumber
&\phantom{=}&\qquad \left.+ a_{33}(i)\rund{ \mathcal{K}_{ij} \mathcal{K}_{ik}} \right] \\
\nonumber
&+&\sum_{m=1}^{N_{\rm
    M}}b_{11}(m)\mathcal{D}^{(1)}_{mj}\mathcal{D}^{(1)}_{mk} +
b_{22}(m)\mathcal{D}^{(2)}_{mj}\mathcal{D}^{(2)}_{mk} \\
&+& \eta \sum_{g} \mathcal{K}_{gj} \mathcal{K}_{gk}\; , 
\label{eq:A3}
\end{eqnarray}
where the sums over $i$, $g$ and $m$ denote summation over all
galaxies with ellipticity measurements, all grid points, and all images
in the multiple imaged system, respectively. The $V_j$
vector carries the information of all constant terms in \eqref{eq:A1}
\begin{eqnarray}
\nonumber
V_j &=& \sum_{i=1}^{N_{\rm gal}}a_{1}(i)\mathcal{G}^{(1)}_{ij}  +
a_{2}(i)\; \mathcal{G}^{(2)}_{ij}  +   a_{3}(i)\;\mathcal{K}_{ij}  \\
\nonumber
&+&\eta \sum_{g}\kappa^{(n_2-1)} \mathcal{K}_{gj} \\
&+&\sum_{m=1}^{N_{\rm M}} b_{1}(m)\mathcal{D}^{(1)}_{mj} +
b_{2}(m)\;\mathcal{D}^{(2)}_{mj} \; .  
\label{eq:A4}
\end{eqnarray}

The coefficients $a$ now differ depending on whether we are in the $\abs{g} \le
1$ or $\abs{g} > 1$ regime. For $\abs{g} \le 1$ the coefficients are given by
\begin{eqnarray}
\nonumber
  a_{11}(i) &=& a_{22}(i) = \frac{Z^2}{\hat{\sigma}_{\le}^2} \; ;\;\;\; 
  a_{13}(i) =  \frac{Z^2}{\hat{\sigma}_{\le}^2} \epsilon_{1} \; ; \\ 
\nonumber
 a_{23}(i) &=&  \frac{Z^2}{\hat{\sigma}_{\le}^2} \epsilon_{2} \; ;\;\;\;
 a_{33}(i) =  \frac{Z^2}{\hat{\sigma}_{\le}^2} \abs{\epsilon}^2 \; ;\\
  a_{1}(i) &=&  \frac{Z}{\hat{\sigma}_{\le}^2} \epsilon_{1} \; ; \; 
  a_{2}(i) =  \frac{Z}{\hat{\sigma}_{\le}^2} \epsilon_{2} \; ; \; 
 a_{3}(i) =  \frac{Z}{\hat{\sigma}_{\le}^2} \abs{\epsilon}^2 \;.   
\label{eq:A5}
\end{eqnarray}
For $\abs{g}>1$ case we get
\begin{eqnarray}
\nonumber
  a_{11}(i) &=& a_{22}(i) = \frac{Z^2}{\hat{\sigma}_{>}^2}\abs{\epsilon}^2\; ;\; \;\; 
  a_{13}(i) =  \frac{Z^2}{\hat{\sigma}_{>}^2} \epsilon_{1} \; ; \\
\nonumber 
 a_{23}(i)& =&  \frac{Z^2}{\hat{\sigma}_{>}^2} \epsilon_{2} \; ; \;\;\;  
 a_{33}(i) =  \frac{Z^2}{\hat{\sigma}_{>}^2} \; ; \\
  a_{1}(i) &= & \frac{Z}{\hat{\sigma}_{>}^2} \epsilon_{1} \; ; \; 
  a_{2}(i) =  \frac{Z}{\hat{\sigma}_{>}^2} \epsilon_{2} \; ; \; 
 a_{3}(i) =  \frac{Z}{\hat{\sigma}_{>}^2} \; .  
\label{eq:A6}
\end{eqnarray}

The coefficients $b$ are carrying the information
about the multiple imaged system:
\begin{eqnarray}
\nonumber
  b_{11}(m) &=& \frac{Z^2(z_{\rm s})}{\sigma_{{\rm s},m(1)}^2}\; ;\; \;\; 
  b_{22}(m) =  \frac{Z^2(z_{\rm s})}{\sigma_{{\rm s},m(2)}^2}  \; ; \\
\nonumber  
 b_{1}(m) &=& \frac{Z(z_{\rm s})\rund{\theta_{m,1}-\beta_{{\rm
  s},1}}}{\sigma_{{\rm s},m(1)}^2} \; ;  \\
  b_{2}(m) &=&  \frac{Z(z_{\rm s})\rund{\theta_{m,2}-\beta_{{\rm s},2}}}{\sigma_{{\rm s},m(2)}^2}  \; ,
 \label{eq:A7}
\end{eqnarray}
where  $\theta_{m,1}$, $\theta_{m,2}$, and $\sigma_{{\rm s},m(1,2)}$
are defined in Sect.~\ref{sc:chi2}. For the reasons mentioned in
Sect.~\ref{sc:chi2} we consider the measurement errors
projected to the source plane isotropic, we set
$\sigma_{{\rm s},m(1,2)}$ equal for the purpose of the reconstruction.

\bibliography{/home/marusa/latex/inputs/bibliogr_clusters,/home/marusa/latex/inputs/bibliogr,/home/marusa/latex/inputs/bibliogr_cv}
\bibliographystyle{/home/marusa/latex/inputs/aa}
\end{document}